\documentclass[12pt]{article}

\begin{document}

\bigskip

\smallskip
\begin{center}                                
    CALCULATION OF PSEUDOSCALAR AND VECTOR MESONS' MASSES IN EXTENDED
                 MODEL OF QUASI-INDEPENDENT QUARKS

\medskip

                         {\it V.V. Khruschov}
\medskip

        {\small   Russian Research Centre "Kurchatov Institute"}
                         
\end{center}

\smallskip
                            
\begin{abstract}                                
Masses of pseudoscalar and vector mesons, which are $\bar qq'$ ground
states of light or/and heavy quarks and antiquarks, have been
calculated in the framework of an extended model of 
quasi-independent quarks with absolute uncertainties about $30$ MeV. It is
shown that the assumption on equidistant discrete levels for a mean
field energy does not contradict to existing data and permits to
reduce a number of model parameters. It is obtained, that a
difference between  neighbouring mean field energy levels is
flavour independent. Values of spin-spin interaction between quarks
and antiquarks with different flavours in pseudoscalar and
vector mesons are presented.   

\end{abstract}

\vspace*{0.5mm}

  At the present time an identification of hadrons bearing
additional constituents in comparison with standard ones
is the topical problem for strong interaction physics. For this 
purpose a detailed description of characteristics of standard hadrons 
is needed. However, in the QCD framework this problem has not been 
solved in the full extent yet. For instance,
although a considerable
progress in Lattice QCD evaluations was achieved the evaluation of
hadron masses is not possible with precision compatible with
existing data accuracies \cite{yao}. Especially it concerns the
evaluation of hadron masses for ground and exited states with
light quarks and/or antiquarks. Hence phenomenological models,
which do not contradict QCD principles, retain their validity. One
of the most interesting and simplest hadron models is the
relativistic model for quasi-independent quarks. The main
statement of this model is the following: hadron's properties can be
described considering a hadron as a system of independent
constituents (or quasi-independent ones with weak residual
interactions), which move in some mean self-consistent field \cite{bog}.
For a description of constituent's motion the one-particle
equation (Schr\"odinger, Dirac or Klein-Gordon-Fock equation) can be
used. The confinement of color particle is taken into account
by using of a linear rising potential, for which the hypothesis of its flavour
independence has been proposed \cite{doro,mart,quroth}.

  The relativistic model for quasi-independent quarks has been
applied with the particular potential entered in the Dirac
equation for a description of meson properties \cite{khrusas,kh}.
The potential used is a generalization of the Cornell potential and
consist of the relativistic vector quasi-Coulomb potential and the
relativistic scalar linear rising potential. In the framework of
this model the hypothesis of universality of confinement potential
offered in Refs.\cite{doro,mart,quroth} has been confirmed for heavy and light
quarks. The coefficient of linear rising scalar potential (string
tension) has been found to be $\sigma  = 0.20\pm 0.01 \, GeV^2$. It should be
noted that in order to describe jointly the ground states of
pseudoscalar and vector mesons different values for mean field
energy have been used. In the present paper we propose that these
values are not simply some phenomenological constants, but
represent in a rough approximation equidistant levels for mean
field energy, which do not depend on quark flavours. We find the
values of energy levels more precisely in combination with
magnitudes of spin-spin interaction for quarks and antiquarks with
different flavours inside the pseudoscalar and vector mesons.

  In the main approximation of the model of quasi-independent
quarks, when one neglect the so-called residual interactions,
which are supposed to be weak, the energy of the composed system
is equal to the sum of constituents' energies. We include the mean
self-consistent field as a quasi-classical object in a set of
constituents, so the hadron energy $E_h$ is
\begin{equation}
E_h=E_0+E _1+...+E _n,
\end{equation}
where $E_0$ is a mean field energy, $E_i$ is a energy of $i-$th
constituent, $i = 1, 2, ... , n$. However, in order to improve the
accuracy of calculations an account of residual interactions'
contributions is needed. In the following we consider ordinary
(non exotic) hadrons and relate the model of quasi-independent
quarks to the constituent quark model.  It will be obtained that
energies of fermionic constituents $E_i$ (quark mass terms) for meson
ground states are equal to constituent quarks' masses within
uncertainties of model of quasi-independent quarks. Some
complications arise when one treats exited states. In the model of
quasi-independent quarks $E_i$ remains an exited energy of $i-$th
constituent, while in the constituent quark model besides
constituent quarks' masses an additional contribution representing
energy of excitation must be taken into consideration.

  For the pseudoscalar and vector mesons the spin-spin interaction
may significantly account for a residual interaction in the ground
states of $\bar qq'$ system within the accuracy of the model. Therefore
mesons' masses can be evaluated by means of the following formula:
\begin{equation}
M_{m}=E_{0m}+E_{1}
+E_{2}
+4<\mbox{\boldmath
$s_1$}\mbox{\boldmath $s_2$}>_{q_{1}\bar{q}_{2}}V_{ss}^{q_{1}\bar{q}_{2}}
\label{me}
\end{equation}
\noindent where $\mbox{\boldmath $s_1$}$/$\mbox{\boldmath $s_2$}$ 
is a quark/antiquark spin,  $V_{ss}^{q_{1}\bar{q}_{2}}$  is a magnitude of a
spin-spin interaction.  It should be noted that in the framework
of the constituent quark model one can find the similar formula
\cite{zeldsa}:
\begin{equation}
 M_m =  m_0 +  m_1 + m_2 + (\mathbf{s}_{1}\mathbf{s}_{2})v_{ss},
\label{fueqq}
\end{equation}
where $m_1$, $m_2$ are  masses of constituent quarks, $m_0$ is some additional
phenomenological contribution.  So we relate constituents'
energies $E_i , i = 1, 2,$ with constituent quarks' masses $m_i, i = 1,
2$. In what follows this assumption will be supported by numerical
fit for meson masses. Note that constituent quarks can be applied
successfully in hadron spectroscopy, as well as  for description
of effects in hadron collisions (see, e.g. \cite{tt}).

  At present due to experimental uncertainties and mixing effects
there is no only one solution for the set of parameters entering
in Eq.(\ref{me}). For instance, there is a well known difficulty related
to the mass value of the $\pi-$meson. Below we give an admissible fit
with minimal number of parameters for the mass values of
pseudoscalar and vector mesons, which can be considered as
definite $\bar qq'$ systems involving $u-, d-, s-, c-, b-$quarks and
antiquarks.

  To do this, we use in an essential manner the results obtained
in Refs.\cite{khrusas,kh}. In these papers in order to determinate $E_i$ for
the $i-$th quark (antiquark) moving in the mean self consistent
field inside the meson the Dirac equation with a static QCD
motivated potential $V(r)$ was used. The potential $V(r)$ had been
chosen in the spherically symmetrical form as a sum of a Lorentz
scalar part $V_0(r)$ and a Lorentz vector part $V_1(r):$$ V(r) =  \beta V_0(r) +
V_1(r)$. The equation for a quark or antiquark with current mass $m_i$
can be written in the following form:
\begin{equation}
\sqrt{\lambda _{i}+m_{i}^{2}}\psi _{i}(\mathbf{r_{i}})=\left[ (%
\mbox{\boldmath
$\alpha_i$}\mathbf{p_{i}})+\beta(m_{i}+V_{0})+V_{1}\right] \psi _{i}(%
\mathbf{r_{i}}),
\label{diq}
\end{equation}
with $\lambda_i$ is an eigenvalue of a radial equation, 
$E_{i}(n_{i}^{r},j_{i})=$ $\sqrt{\lambda _{i}+m_{i}^{2}}$,
 $i=1,2$,  $V_{0}(r)=$ $%
\sigma r/2$ and $V_{1}(r)=$ $-2\alpha _{s}/3r$. Here, $n_i^r$ and $j_i$
  are a radial
quantum number and a quantum number of an angular moment for an $i-$th 
constituent, model parameters  $\sigma $ and $\alpha _{s}$ have meanings 
of the 
coefficient of the linear rising potential (string tension) and the
strong coupling constant at small distances, correspondingly.

In  spite of the fact that the quark and antiquark energies $E_1$
and $E_2$ bring in  main contributions to meson masses in order to
reduce relative uncertainties of evaluations up to $10^{-2}$ level for
all mesons the additional contributions due to the $E_0$ and $V_{ss}$ are
needed. Along similar lines $E_0$ and $V_{ss}$ contributions were
considered in certain of approaches \cite{zeldsa,fili}, however
some ambiguities occur when $E_0$ and $V_{ss}$ are evaluated. When
different fits for meson masses had been performed the case was
adopted, such that the $V_{ss}$ value for $u-,d-$quarks and antiquarks
was equal to $100 MeV$ and the $E_0$ value for $\rho-$meson was zero. In this
case the $E_0$ value for $\pi -$meson is double that the $E_0$ value for $K$
meson: $E_0^{\pi} = 2E_0^K$, where $E_0$ values are marked at the top with
mesons' names. The nonzero values of $E_0$ and $V_{ss}$ given in MeV units
for different pseudoscalar and vector mesons are listed below.
\begin{eqnarray}
  E_0^K = -118,  E_0^{J/\psi} = -125,  E_0^{\eta_c} = -230, 
E_0^{\eta_b} = -600,  \nonumber\\
E_0^{\Upsilon} = -450,  E_0^{D_s} = -112,   E_0^{B_c} = -270, V_{ss}^{K^*} = 70, 
V_{ss}^{\phi} = 50, \\
V_{ss}^{D^*} = 27,  V_{ss}^{D_s^*} = 5,   V_{ss}^{J/\psi} = 3, V_{ss}^{B^*} = 3, 
V_{ss}^{B_s^*} = 2, V_{ss}^{D_c^*} = 1.3,V_{ss}^{\Upsilon}
= 1\nonumber
\end{eqnarray}
  The uncertainties of determination of the $E_0$ and $V_{ss}$ values are
of the order of several MeV. As is seen in the list of the $E_0$
values above all nonzero $E_0$ values for ground states of
pseudoscalar and vector mesons for all quark flavours are
approximately multiples to the $E_0^K$. Thus we can accept that the $E_0$
values  are varied according to the rule:
\begin{equation}
                        E_0^n = -708 + n E_0^K,
\end{equation}                                
where n is a negative integer. To give also the values of
constituents' energies $E_i$ for $u-, d-, s-, c-, b-$quarks and
antiquarks (for  two super light $u-$ and $d-$quarks and antiquarks
the mean value $E_N$ is presented):
\begin{equation}
  E_N = 337\pm 3,  E_S = 485\pm 8, E_C = 1610\pm 15, E_B = 4952\pm
20  
\label{const1}
\end{equation}
Notice that within the model uncertainties the constituents'
energies for light quarks and antiquarks agree with the
constituent quark masses obtained in Refs.\cite{del,sca}.

  The model of quasi-independent quarks supplemented by the
proposition about the equidistant discrete levels of mean field
may be named the extended model of quasi-independent quarks. In
the framework of this model it is possible to evaluate the
pseudoscalar and vector mesons' masses with relative uncertainties of order
$10^{-2}$. From the
results presented it may be inferred that the extended model of
quasi-independent quarks is a workable generalization of the constituent
quarks model. 
It is highly plausible that the formation of
constituent quarks take place in the nonperturbative region with
the characteristic scales, which can be evaluated with the
universal coefficient of a slope of a linear rising potential $\sigma =
0.20\pm 0.01 \, GeV^2$. The scales' values with dimensions of mass and
length are equal to $\mu_C = 0.45\pm0.02 \, GeV$, 
$\lambda_C = 0.44\pm0.02 \, Fm$.  As this
takes place, $\mu_C$ defines typical magnitudes of transversal impulses
$<p_T>$ for quarks-partons inside hadrons, while a radius of a
perturbative region surrounded a current quark is equal to $r_C =
\lambda_C/2 = 0.22\pm0.01 \, Fm$. The region $r > r_C$  is most likely to 
be the region of a formation for a constituent quark due to
nonperturbative interactions.  If one take into account typical sizes
of hadrons, then a radius of a constituent quark is $0.25 \div 0.35 \, Fm$.
The absolute value of  $E_0^K$, which
determinates the characteristic scale for mean field energies in
the framework of the extended model of quasi-independent quarks
can also be expressed in terms of $\mu_C$ as $\mu_C/4$.

     The  author  is grateful to Yu.V.  Gaponov,  V.I.
Savrin, S.V. Semenov and A.M. Snigirev for useful discussions. The
work  was  partially supported by the grant   26  on  fundamental
researches of the RRC "Kurchatov Institute" in 2007 year.

\end{document}